\documentclass[prc,twocolumn,showpacs,preprintnumbers,amsmath,amssymb]{revtex4}
\usepackage{graphicx}% Include figure files
\usepackage{dcolumn}% Align table columns on decimal point
\usepackage{bm}% bold math
\def\be{\begin{equation}}
\def\ee{\end{equation}}
\def\ba{\begin{array}}
\def\ea{\end{array}}
\def\bea{\begin{eqnarray}}
\def\eea{\end{eqnarray}}
\begin{document}

\title{Decay analysis of compound nuclei with mass A$\sim 30-200$ \\ formed in the reactions involving loosely bound projectiles\\}

\author{Mandeep Kaur$^1$, BirBikram Singh$^{1*}$, Manoj K.Sharma$^2$ and Raj K.Gupta$^3$}

\affiliation{$^1$Department of Physics, Sri Guru Granth Sahib World University,
Fatehgarh Sahib-140406, India.\\
$^2$SPMS, Thapar University,
Patiala-147004, India.\\
$^3$Department of Physics, Panjab University,Chandigarh-160014, India.}

\date{\today}
\begin{abstract}
The dynamics of the reactions forming compound nuclei using loosely bound projectiles is analysed within the framework of dynamical cluster decay model (DCM) of Gupta and Collaborators. We have analysed different reactions with $^{7}Li$, $^{9}Be$ and $^{7}Be$ as neutron rich and neutron deficient projectiles, respectively, on different targets at the three $E_{lab}$ values, forming compound nuclei within the mass region A$\sim 30-200$. The contributions of light particles LPs ($A\le4$) cross sections $\sigma_{LP}$, energetically favoured intermediate mass fragments IMFs ($5 \le A_2 \le 20$) cross sections $\sigma_{IMF}$ as well as fusion-fission $\it{ff}$ cross sections $\sigma_{ff}$ constitute the $\sigma_{fus}$ (=$\sigma_{LP}$+$\sigma_{IMF}$+$\sigma_{ff}$) for these reactions. The contribution of the emitted LPs, IMFs and ff fragments is added for all the angular momentum upto the $\ell_{max}$ value, for the resepctive reactions. Interestingly, we find that the $\Delta R^{emp}$, the only parameter of model and uniquely fixed to address the $\sigma_{fus}$ for all other reactions having $\it same$ loosely bound projectile at the chosen incident energy. It may be noted that the dynamical collective mass motion of preformed LPs, IMFs and ff fragments or clusters through the modified interaction potential barrier are treated on parallel footing. We see that the values of modified interaction barrier heights $\Delta V_{B}^{emp}$ for such reactions are almost of the same amount specifically at the respective $\ell_{max}$ values.
\end{abstract}
%\pacs{25.70.Jj, 23.70.+j, 24.10.-i, 23.60.+e}

\maketitle

\section{INTRODUCTION}
The low energy heavy ion reactions forming compound nuclear systems provide unique platform to study the several nuclear properties investigated during last few  decades. Since the advent of heavy ion beam accelerator technology and the latest development of radioactive ion beam (RIB) facilities, a variety of nuclear reactions have been studied or are being studied world over. As a result, lot of experiential data is available to analyse the atomic nucleus and subsequent dynamical behavior, which is still a continuous source of novel and useful phenomenon since its pioneering discovery in 1911. The nuclear reactions at low energy provide opportunity to study not only the structural aspect of the nucleus but also the collective behaviour of its constituents along with adequate inculcation  of temperature and angular momentum effects. Number of reactions involving loosely bound projectiles have been studied or are being further investigated for better understanding of nuclear behavior\cite{Bodek80,Beck03,Marti05,Kalita06,Gomes06,Sinha08,Parkar10,Martinez14,Fang15}. During this time enhancement/ suppression of fusion cross sections around the Coulomb barrier have been observed and lot of experiential data including fusion and breakup cross sections for such reactions is available. Theoretical descriptions of these observations have been exercised simultaneously, but still need further emphasis. Recently a systematic study of reactions induced by loosely bound projectiles have been made \cite{Singorini97,Phookan13}. However, the fast development in the nuclear reaction technology has outpaced the deeper theoretical analysis of the experimental data.

To meet the above mentioned challenge Gupta and Collabotators are developing the dynamical cluster decay model (DCM) to study the heavy ion reactions at low energies. The DCM, for the decay of compound systems, is non-statistical description of dynamical mass motion of preformed clusters through the interaction barrier which treats all types of emissions i.e. evaporation residues (or equivalently light particles LPs ($A\le4$)), intermediate mass fragments IMFs ($5 \le A_2 \le 20$) and fusion-fission $\it{ff}$ fragments, on the same footing \cite{Gupta05,Gupta06,Singh06,Singh08,Sharma09, Gupta09,Kumar09,Gupta10,Gupta11,Sharma12,Kaur12,Sandhu12,Bansal12,Kaur13,Niyti14,Kaur14}. But the statistical models treat all these emissions on different footing, where these emissions are treated differently on the basis of mass of compound nucleus\cite{Morreto75,Campo91,Matsus97}. During last decade number of reactions in the very light, light, medium, heavy and superheavy mass region have been studied successfully for different type of decays characterizing the particular mass region of the compound nucleus. The exclusive studies have been made for the effect of nuclear structure, shape (deformations and orientations), temperature and angular momentum on such decays. It is relevant to mention here that the neck length parameter $\Delta R^{emp}$ is the only parameter of this model, which is fixed empirically. Gupta and Collaborators successfully presented interaction barrier modification characteristic of $\Delta R$ in significant studies \cite{Gupta09, Kumar09, Gupta10, Gupta11, Bansal12}. Interestingly, these works point out that the empirically fitted $\Delta R^{emp}$ simply result in the corresponding 'barrier lowering' $\Delta V_{B}^{emp}$ for the given reaction.

 In the present work, the study of the reactions around the coulomb barrier involving loosely bound projectiles $^{7}Li$, $^{7}Be$ and $^{9}Be$ on the different targets have been undertaken.The experimental data for the reactions $^{7}Be$ + $^{27}Al$ \cite{Kalita06}, $^{7}Be$ + $^{58}Ni$\cite{Martinez14}, $^{7}Li$ + $^{27}Al$ \cite{Kalita06}, $^{7}Li$ + $^{28}Si$ \cite{Sinha08}, $^{7}Li$ + $^{59}Co$ \cite{Beck03}, $^{9}Be$ + $^{27}Al$ \cite{Marti05}, $^{9}Be$ + $^{28}Si$ \cite{Bodek80}, $^{9}Be$ + $^{124}Sn$ \cite{Parkar10}, $^{9}Be$ + $^{144}Sm$ \cite{Gomes06}, $^{9}Be$ + $^{169}Tm$ and $^{9}Be$ + $^{187}Re$ \cite{Fang15} is available for the fusion cross-section. We have studied the dynamics of these reactions within the framework of DCM. In addition, we have investigated some other reactions with the same set of projectiles and energies on stable targets. The reactions involving $^{7}Li$ and $^{7}Be$ projectiles on stable targets such as $^{32}S$, $^{40}Ca$, $^{48}Ti$ have been studied. In addition, $^{7}Be$ on $^{65}Cu$ have also been studied. It is interesting to note that the value of $\Delta R^{emp}$ is fixed empirically for a particular projectile at a given energy and the fusion cross-sections of all the reactions for that projectile is calculated at the same value. Using the $\Delta R^{emp}$ for a given projectile, predictions are made for the fusion cross sections of a reaction for which experimental data is not available.

The main aim of the present work is to bring forth, further, the significance of neck length parameter $\Delta R^{emp}$ which simply amounts the interaction potential barrier modification for the preformed clusters, around the Coulomb barrier. This study is further attempt to establishing the predictability of the DCM. So, we attempted to study the fusion cross sections $\sigma_{fus}$, of about 18 reactions. The main emphasis is to account a comprehensive addressal of decay mechanism of loosely bound reactions within the framework of collective clusterisation of DCM description. We have confined ourselves to TF process and other aspects such as breakup and incomplete fusion etc may be taken up in near future.

The Section II gives brief description of the DCM for hot and rotating compound nucleus. The calculations and discussions are presented in section III. A summary of results constitute section IV.

\section{THE DYNAMICAL CLUSTER-DECAY MODEL FOR HOT AND ROTATING COMPOUND SYSTEMS}

The DCM of Gupta and collabrators  is worked out in terms of the collective coordinates of mass (and charge) asymmetries $\eta_A={{(A_1-A_2)}/{(A_1+A_2)}}$ where $A_1$ and $A_2$ are the masses of incoming nuclei. For charge distributions, a corresponding charge-asymmetry coordinates  $\eta _Z={{(Z_1-Z_2)}/{(Z_1+Z_2)}}$ where $Z_1$ and $Z_2$ are the masses of incoming nuclei
and  relative separation R ($\geq R_1 + R_2$), to which is added the multipole deformations $\beta_{\lambda i}$ ($\lambda = 2,3,4$), orientations $\theta i$ (i=1,2) of two nuclei or fragments, and the azimuthal angle $\phi$ between their principal planes.\cite{Gupta73,Maruhn74,Gupta75}\\

In terms of above coordinates, for $\ell$-partial waves, the compound nucleus (CN) decay or the fragment production cross section for each process is
\be
\sigma=\sum_{\ell=0}^{\ell_{max}}\sigma_{\ell}={\pi \over
k^2}\sum_{\ell=0}^{\ell_{max}}(2\ell+1)P_0P; \quad k=\sqrt{2\mu
E_{c.m.}\over {\hbar^2}} \label{eq:1} \ee

where the preformation probability $P_0$ refers to $\eta$ motion and the penetrability
$P$ to R motion, both depending on angular momentum $\ell$ and temperature T, because the CN excitation energy $E_{CN}^*=E_{c.m.}+Q_{in}$=$1\over{9}$ $A_{CN}T^2-T$; $Q_{in}$ is the entrance channel Q-value. The deformation of nuclei are kept fixed in R-motion, and independent of temperature T. $\ell_{max}$ is the maximum angular momentum, fixed for the vanishing of the light particles cross-section, i.e., $\sigma_{ER}$
becoming negligibly small at $\ell=\ell_{max}$. Then total decay cross-section is given by
\be
\sigma_{fus}= \sigma_{LP}+\sigma_{IMF}+\sigma_{ff}+\sigma_{nCN}
\label{eq:2}
\ee
The preformation probability $P_0$  is given by
$P_0=\mid \psi_R(\eta (A_i))\mid ^2 {\sqrt{B_{\eta \eta }}}{2\over A_{CN}}$ is obtained by solving the stationary Schr\"odinger equation in $\eta$, at a fixed $R=R_a$,
\be
\{ -{{\hbar^2}\over {2\sqrt B_{\eta \eta}}}{\partial \over {\partial
\eta}}{1\over {\sqrt B_{\eta \eta}}}{\partial\over {\partial \eta
}}+V(\eta ,\beta_{\lambda_i},\theta_i,T)\} \psi ^{\nu}(\eta ) = E^{\nu} \psi ^{\nu}(\eta )\}
\ee
with $R_a=R_1(\alpha_1,T)+R_2(\alpha_2,T)+{\Delta R}(T)$. The mass parameters $B_{\eta\eta}$ are the smooth hydrodynamical
masses \cite{Kroger80}. The radius vector is given by
\be
R_i(\alpha_i,T)=R_{0i}(T)\Big[1+\sum_{\lambda}\beta_{\lambda i}Y_{\lambda}^{(0)}(\alpha_i)\Big]
\ee
$R_{0i}(T)$ are temperature dependent nuclear radii for equivalent spherical nuclei \cite{Royer92}.
\be
R_{0i}(T)=[1.28A_i^{1/3}-0.76+0.8A_i^{-1/3}](1+0.0007 T^2)
\ee
In Eq.(3), $\nu$ = 0,1,2,3,.....refer to ground state ($\nu$=0) and excited state solutions and, for a Boltzmann-like function,$|\psi|^2$ = $\sum_{\nu=0}^{\infty}\beta_|\psi^{(\nu)}|^2\exp(-E^{(\nu)/T})$. $R_a$ is the first turning point of penetration path used for calculating the penetrability P.\\

The angle $\alpha_i$ in the above equation is that which the nuclear symmetry axis makes with the radius vector $R_i(\alpha_{i}$, measured in clockwise direcrtion, see Fig.1 in Ref.\cite{Kaur14}. This is to be distinguishing from the orientation angle $\theta_i$  that the nuclear symmetry axis makes with the collision Z axis, measured in anticlockwise direction. In the language of the two center shell model (TCSM) used to determine the shell effect $\delta U$, $\Delta R$ is shown to assimilate the neck formation effects, hence referred to as a neck length parameter. As we have pointed out earlier that $\Delta R^{emp}$ has been fitted empirically for a reaction having experimental $\sigma_{fus}$ given for a projectile, while $\sigma_{fus}$ for other reactions (with the same projectile) have been either predicted or compared with the available experiential data.

\begin{figure}[t]
\includegraphics[width=0.85\columnwidth,clip=true]{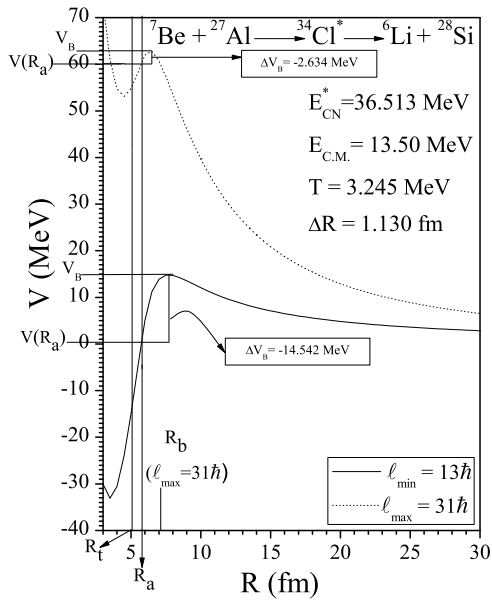}
\vspace{-0.8cm}
\caption{The Scattering potentials $V(MeV)$ for the compound system $^{34}$Cl$^*$ (formed in the reaction $^{7}$Be+$^{27}$Al) for exit channel $^{6}$Li+$^{28}$Si at the two extreme $\ell$-values. }
\label{Fig.1}
\end{figure}

\begin{table*}
\caption{\label{tab:table1}The calculated barrier modification factor $\Delta V_{B}$ = $ V({R_a})$ - $V_{B}$ at different $\ell$-values for the interaction potential for $^{7}$Be induced reactions at $E_{lab}$$ \sim 17$ MeV.  }
\begin{ruledtabular}
\begin{tabular}{llllll}
\cline{1-6}
&&&\multicolumn{3}{c}{$\Delta V_{B}$ (MeV)} \\
Reaction &$\ell_{min}$ ($\hbar$)&$\ell_{max}$ ($\hbar$)&$\ell_{min}$ ($\hbar$)&$\ell$ = 20($\hbar$)&$\ell_{max}$ ($\hbar$)\\
\hline\\
$^{7}$Be+$^{27}$Al$\rightarrow$$^{34}Cl^*$$\rightarrow$$^{6}$Li+$^{28}$Si&13&31&-14.542&-3.616&-2.634\\
$^{7}$Be+$^{32}$S$\rightarrow$$^{39}Ca^*$$\rightarrow$$^{6}$Li+$^{33}$Cl&8&29&-9.065&-4.789&-2.005\\
$^{7}$Be+$^{40}$Ca$\rightarrow$$^{47}Cr^*$$\rightarrow$$^{6}$Li+$^{41}$Sc&4&30&-9.102&-5.449&-2.720\\
$^{7}$Be+$^{48}$Ti$\rightarrow$$^{55}Fe^*$$\rightarrow$$^{6}$Li+$^{49}$V&5&36&-9.317&-7.191&-2.804\\
$^{7}$Be+$^{58}$Ni$\rightarrow$$^{65}Ge^*$$\rightarrow$$^{6}$Li+$^{59}$Cu&0&38&-9.984&-7.180&-2.970\\
$^{7}$Be+$^{65}$Cu$\rightarrow$$^{72}As^*$$\rightarrow$$^{6}$Li+$^{66}$Zn&0&44&-10.715&-9.134&-3.970\\
\end{tabular}
\end{ruledtabular}
\end{table*}

The parameter $\Delta R$ fixes the first turning point of the barrier penetration, refereing to the actually used barrier height V($R_a$) consequently to the concept of barrier lowering $\Delta V_B$($\ell$). The choice of $\Delta R$ for the best fit to the data allows us to define the effective "barrier lowering" parameter $\Delta V_B$ for each $\ell$ as the difference between the actual used barrier $V(R_a,\ell)$, as

\be
\Delta V_{B} = V(R_{a}, \ell) - V_B(\ell)
\ee
$\Delta V_{B}$ values at two extreme $\ell$-values are defined as the negative quantity as shown in Fig. 1 and hence the actually used barrier is effectively lowered.

In Table. 1 we have presented barrier modification factor $\Delta V_B$ which is defined in Eq. (6) for $^{7}Be$ induced reactions. We observe that the barrier modification or lowering in barrier height is almost of the same amount for $^{7}Be$ induced reactions at $\ell_{min}$ and $\ell$ = 20$\hbar$, but at $\ell$ = $\ell_{max}$ the barrier modification factor is almost constant. It means that almost same amount of modification in the barrier takes place in those reactions which are induced by the same projectile having same incident energy. Subsequently, we can fit the fusion cross-section of the reactions induced by same projectile having same value of projectile energy with same value of empirically fitted neck length parameter $\Delta R^{emp}$.
\begin{figure}[t]
\includegraphics[width=0.85\columnwidth,clip=true]{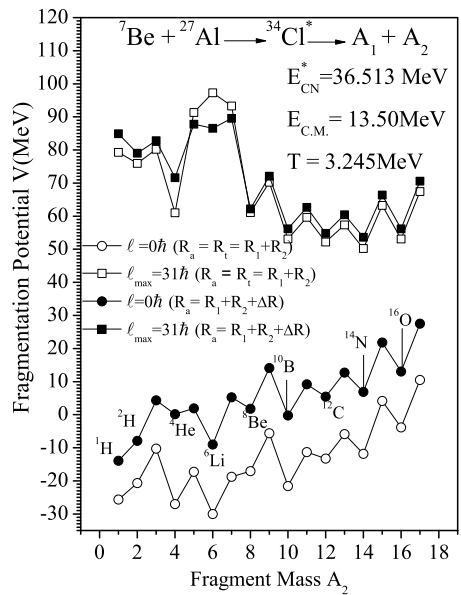}
\vspace{-0.8cm}
\caption{The fragmentation potentials $V(MeV)$ for the compound system
$^{34}$Cl$^*$ formed in the reaction $^{7}$Be+$^{27}$Al for different $\ell$-values at different choices of $\Delta R\neq 0$ and $\Delta R = 0$.}
\label{Fig. 2}
\end{figure}

\begin{figure}[t]
\includegraphics[width=0.85\columnwidth,clip=true]{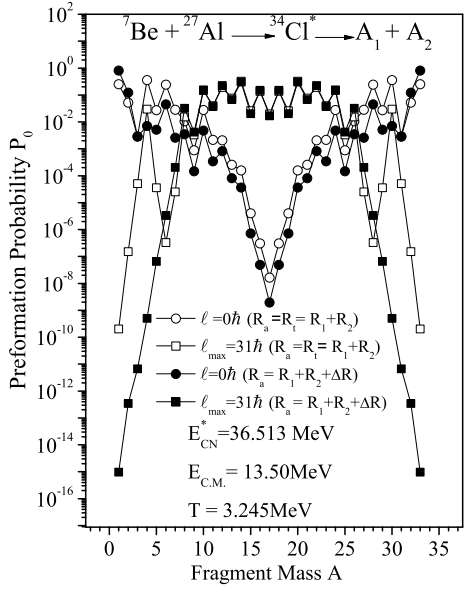}
\vspace{-0.8cm}
\caption{Same as Fig. 2 but for the fragment preformation factor $P_0$. }
\label{Fig.3}
\end{figure}

The structure information of the compound nucleus enters the preformation probability $P_0$ through the fragmentation potential $V_R(\eta, T )$, defined as
\begin{small}
\bea
V_R(\eta,\beta_{\lambda i},\theta_i,T)&=&\sum_{i =1}^{2}[V_{LDM}(A_i,Z_i,T)]+\sum_{i=1}^{2}[\delta U_i]\exp(-T^2/T_0^2)\nonumber\\
&&+V_C(R,Z_i,\beta_{\lambda i},\theta_i,T) +V_P(R,A_i,\beta_{\lambda i},\theta_i,T)\nonumber\\&&+V_{\ell}(R,A_i,\beta_{\lambda i},\theta_i,T).\eea
\end{small}
\par\noindent
Here $V_{LDM}$ and $\delta U$ are, respectively, the T-dependent liquid drop and shell correction energies \cite{Myres80}.

The $V_P$ is an additional attraction due to the nuclear proximity potential \cite{Blocki77}, $V_c$ represents the coloumb potential, $V_{\ell}$ accounts for angular momentum part of interaction.

\begin{figure}[t]
\includegraphics[width=0.85\columnwidth,clip=true]{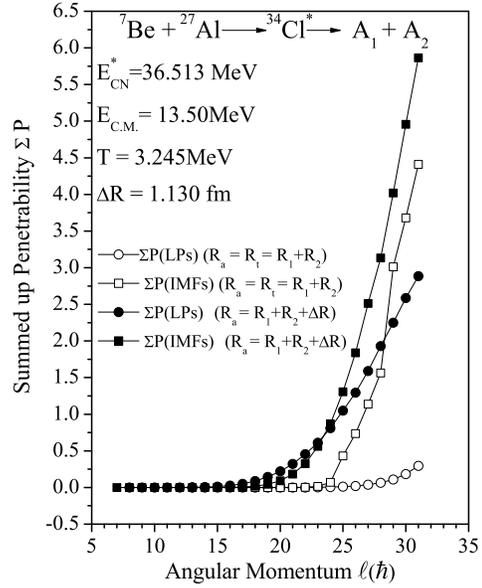}
\vspace{-0.8cm}
\caption{The variation of summed up Preformation Probabilty with angular momentum $\ell$ ($\hbar)$ for $^{34}$Cl$^*$ decay into both LPs and IMFs.}
\label{Fig.4}
\end{figure}

\begin{figure*}[t]
\includegraphics[width=1.3\columnwidth,clip=true]{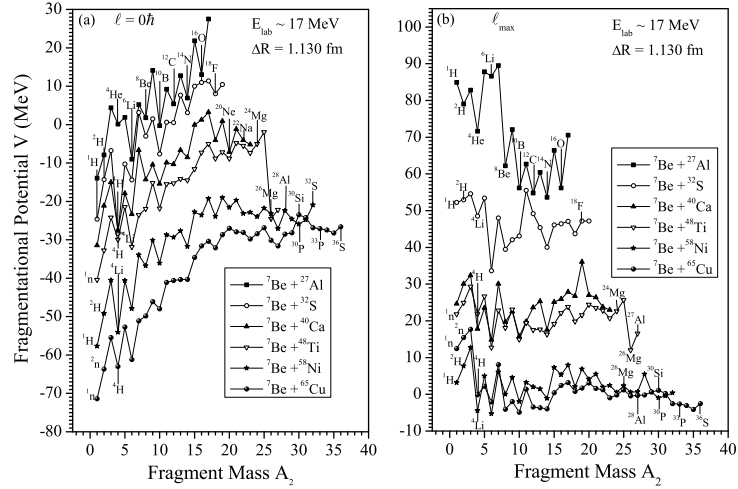}
%\vspace{-2.5cm}
\caption{The fragmentation potentials $V(MeV)$ for the compound systems
having mass A $\sim$ 30-70 formed in $^{7}$Be induced reactions at incident energy $E_{lab}$$\sim$17 MeV  for (a) $\ell$=0$\hbar$ (b)$\ell$=$\ell_{max}$}.
\end{figure*}

\begin{figure*}[t]
\includegraphics[width=1.2\columnwidth,clip=true]{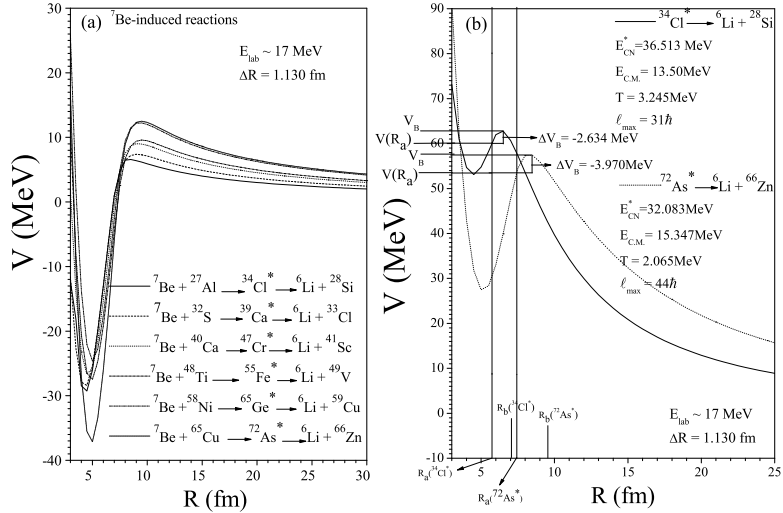}
%\vspace{-2.5cm}
\caption{(a).The barrier height $V_{B}(MeV)$ for the nuclei induced by $^{7}Be$ projectiles at $\ell$=0$\hbar$ values.(b) The first and second turning points of $^{34}$Cl$^*$ and $^{72}$As$^*$ formed in $^{7}$Be+$^{27}$Al and $^{7}$Be+$^{65}$Cu reactions, respectively at $\ell_{min}$ and  $\ell_{max}$.}
\end{figure*}

\begin{figure}[t]
\includegraphics[width=0.85\columnwidth,clip=true]{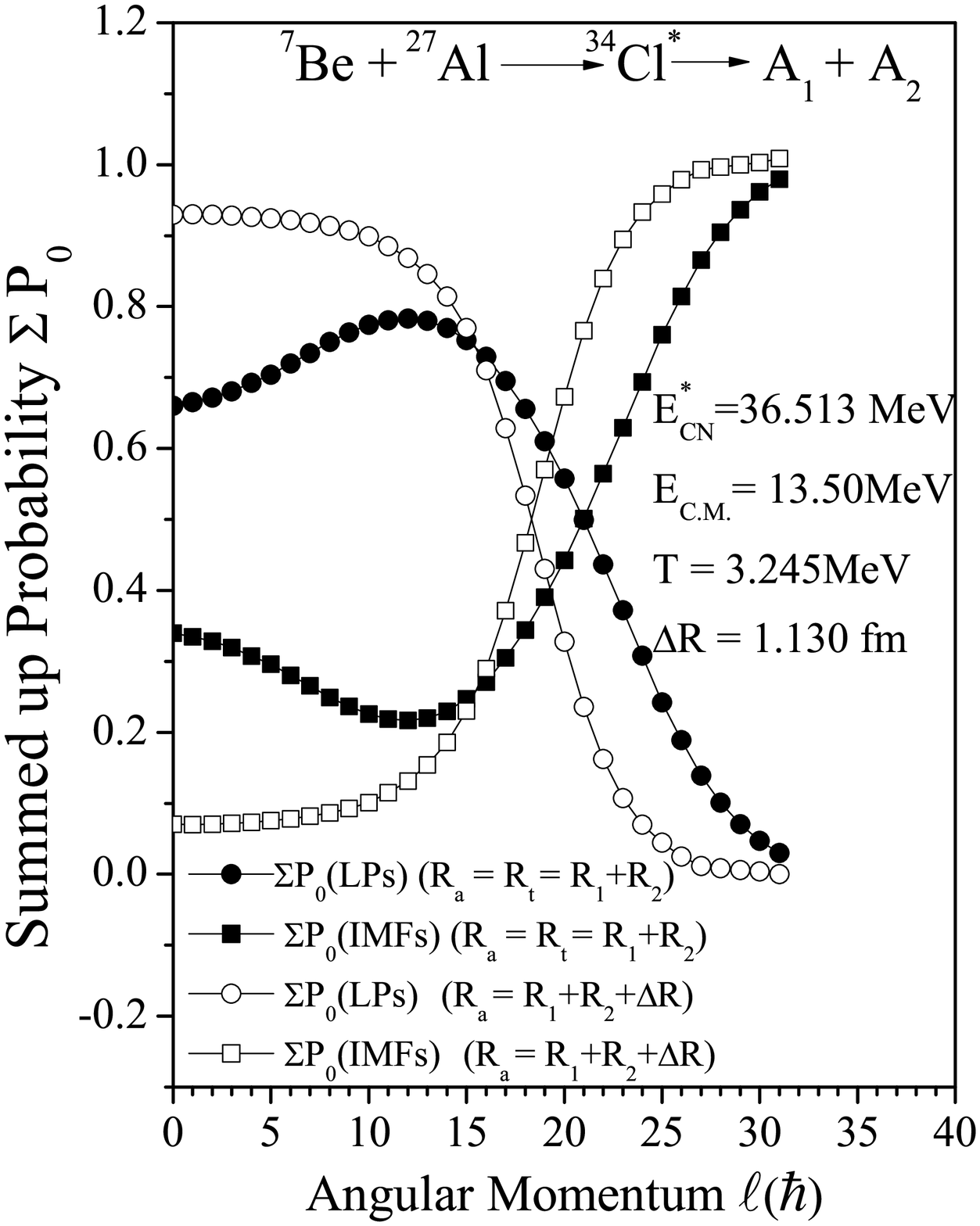}
%\vspace{-1.5cm}
\caption{Same as Fig. 4 but for summed up Penetration Probabilty.}
\label{Fig.7}
\end{figure}

The penetrability calculated as the WKB tunneling probability
\be
P=\exp[{-2\over\hbar}\int_{R_a}^{R_b}\sqrt{2\mu[V(R)-Q_{eff}]}dR]
\ee
$R_a$, defined above, is the first turning point of the penetration path used for calculating the WKB penetrability $P$. It act as a parameter through $\Delta R$($\eta$,T) and we define $R_a$ is same for all $\ell$- values.

\section{CALCULATIONS AND DISCUSSIONS}

The dynamics of the reactions induced by loosely/ weakly bound projectiles $^{7}$Li, $^{7}$Be and $^{9}$Be is studied using DCM around the coloumb barrier energies. The aim of present paper is to address the fusion data of number of loosely bound reaction with specific aim to analyse the characteristic behaviour of neck length parameter $\Delta R$ so that the barrier modification and fusion cross section ($\sigma_{fus}$) may be predicted within confidence and compared nicely with the available experimental data, for all these reactions. Following the neck length systematics, $\sigma_{fus}$ have been predicted for few reactions, where experimental data is not available. The contribution of $\sigma_{LP}$, $\sigma_{IMF}$ and $\sigma_{ff}$ is taken together to calculate the $\sigma_{fus}$ depending upon the decay modes of different nuclei. This study also points out the barrier modification characterisitic of $\Delta R$ for different reactions induced by the same projectile at a given incident energy. In the present work, we have taken three set of reactions induced by loosely bound projectile $^7{Li}$, $^{7}Be$ and $^{9}Be$ at $E_{Lab}$=10, 17 and 28 MeV, respectively on diffferent targets forming compound systems with $A \sim 30-200$. The $\sigma_{fus}$ of a number of reactions is fitted at same value of $\Delta R^{emp}$ for a particular projectile having same incident energy. This same value of $\Delta R^{emp}$ is then used to made the prediction for $\sigma_{fus}$ for reactions with the loosely bound projectiles on some stable targets. In addition to this the role of $\Delta R$, angular momentum and mass of the target has been studied extensively.

\begin{table*}
\caption{\label{tab:table2}The DCM calculated fusion cross-sections $\sigma_{fus}$ for $^{7}$Be induced reactions on different targets at $E_{lab} \sim 17$ MeV are given at $\Delta R$ = 0 and $\Delta R^{emp.}$ = 1.130 fm, alongwith their comparison with the available experimental data. The predicted $\sigma_{fus}$ for some other reactions are also given.}
\addtolength{\tabcolsep}{-3pt}
\begin{ruledtabular}
\begin{tabular}{llllllll}
\cline{1-8}
&&&&&\multicolumn{3}{c}{$\sigma_{fus.}$ (mb)} \\
Reaction &$E_{c.m.}$&$E_{CN}^{*}$ (MeV)&T (MeV)&$\ell_{max}$ ($\hbar$)&$\Delta R$ =0 fm&$\Delta R^{emp.}$
(fm)&Expt.\\
\hline\\
$^{7}$Be+$^{27}$Al$\rightarrow$$^{34}Cl^*$$\rightarrow$$A_{1} + A_{2}$&13.50&36.513&3.245&31&8.003&646.74&$635\pm76$\cite{Kalita06}\\
$^{7}$Be+$^{32}$S$\rightarrow$$^{39}Ca^*$$\rightarrow$$A_{1} + A_{2}$&13.99&30.823&2.790&29&$6.33\ast10^{-2}$&553.00&-\\
$^{7}$Be+$^{40}$Ca$\rightarrow$$^{47}Cr^*$$\rightarrow$$A_{1} + A_{2}$&14.47&29.952&2.493&30&$4.59\ast10^{-5}$&125.90&-\\
$^{7}$Be+$^{48}$Ti$\rightarrow$$^{55}Fe^*$$\rightarrow$$A_{1} + A_{2}$&14.84&45.263&2.804&36&$1.18\ast10^{-5}$&76.05&-\\
$^{7}$Be+$^{58}$Ni$\rightarrow$$^{65}Ge^*$$\rightarrow$$A_{1} + A_{2}$&14.99&27.009&2.004&38&$3.34\ast10^{-6}$&62.84&$61.1\pm6.9$\cite{Martinez14}\\
$^{7}$Be+$^{65}$Cu$\rightarrow$$^{72}As^*$$\rightarrow$$A_{1} + A_{2}$&15.35&32.083&2.065&44&$3.03\ast10^{-5}$&37.00&-\\

\end{tabular}
\end{ruledtabular}
\end{table*}

\begin{table*}
\caption{\label{tab:table3}The same as Table 2, but for $^{7}$Li induced reactions at incident energy $E_{lab} \sim 10$ MeV and for $\Delta R^{emp.}$ = 0.907 fm.}
\addtolength{\tabcolsep}{-10pt}
\begin{ruledtabular}
\begin{tabular}{llllllll}
\cline{1-8}
&&&&&\multicolumn{3}{c}{$\sigma_{fus.}$ (mb)} \\
Reaction &$E_{c.m.}$&$E_{CN}^{*}$ (MeV)&T (MeV)&$\ell_{max}$ ($\hbar$)&$\Delta R^{emp.}$
(fm)&Expt.\\
\hline\\
$^{7}$Li+$^{27}$Al$\rightarrow$$^{34}S^*$$\rightarrow$$A_{1} + A_{2}$&7.941&35.58&3.203&30&437.13&$415\pm67$\cite{Kalita06}\\
$^{7}$Li+$^{28}$Si$\rightarrow$$^{35}Cl^*$$\rightarrow$$A_{1} + A_{2}$&7.79&30.23&2.919&31&352.64&$352.81\pm8$\cite{Sinha08}\\
$^{7}$Li+$^{32}$S$\rightarrow$$^{39}K^*$$\rightarrow$$A_{1} + A_{2}$&8.205&30.90&2.788&30&264.5&-\\
$^{7}$Li+$^{40}$Ca$\rightarrow$$^{47}V^*$$\rightarrow$$A_{1} + A_{2}$&8.51&30.57&2.519&34&97.05&-\\
$^{7}$Li+$^{48}$Ti$\rightarrow$$^{55}Mn^*$$\rightarrow$$A_{1} + A_{2}$&8.72&32.88&2.402&39&48.52&-\\
$^{7}$Li+$^{59}$Co$\rightarrow$$^{66}Zn^*$$\rightarrow$$A_{1} + A_{2}$&9.883&30.74&2.116&44&2.82&3.97\cite{Beck03}\\
\end{tabular}
\end{ruledtabular}
\end{table*}

\begin{small}
\begin{table*}
\addtolength{\tabcolsep}{-10pt}
\caption{\label{tab:table4}The same as Table 2, but for $^{9}$Be induced reactions at incident energy $E_{lab} \sim 28$ MeV and for $\Delta R^{emp.}$ = 1.138 fm.}
\begin{ruledtabular}
\begin{tabular}{llllllll}
\cline{1-8}
&&&&&\multicolumn{3}{c}{$\sigma_{fus.}$ (mb)} \\
Reaction &$E_{c.m.}$&$E_{CN}^{*}$ (MeV)&T (MeV)&$\ell_{max}$ ($\hbar$)&$\Delta R^{emp.}$
(fm)&Expt.\\
\hline\\
$^{9}$Be+$^{27}$Al$\rightarrow$$^{36}Cl^*$$\rightarrow$$A_{1} + A_{2}$&21.00&44.63&3.470&33&1130.823&1191\cite{Marti05}\\
$^{9}$Be+$^{28}$Si$\rightarrow$$^{37}Ar^*$$\rightarrow$$A_{1} + A_{2}$&21.19&33.05&2.960&31&892.69&$945\pm94$\cite{Bodek80}\\
$^{9}$Be+$^{124}$Sn$\rightarrow$$^{133}Xe^*$$\rightarrow$$A_{1} + A_{2}$&26.18&36.94&1.615&103&82&$89.6\pm3.0$\cite{Parkar10}\\
$^{9}$Be+$^{144}$Sm$\rightarrow$$^{153}Dy^*$$\rightarrow$$A_{1} + A_{2}$&26.54&25.07&1.244&104&0.77&$1.72\pm1.02$\cite{Gomes06}\\
$^{9}$Be+$^{169}$Tm$\rightarrow$$^{178}Ta^*$$\rightarrow$$A_{1} + A_{2}$&26.58&27.23&1.200&104&0.182&0.198\cite{Fang15}\\
$^{9}$Be+$^{187}$Re$\rightarrow$$^{196}Au^*$$\rightarrow$$A_{1} + A_{2}$&26.71&29.59&1.188&113&0.0263&0.030\cite{Fang15}\\
\end{tabular}
\end{ruledtabular}
\end{table*}
\end{small}

Fig. 2 illustrates the mass fragmentation potential V(A) at two extreme $\ell$-values for compound system $^{34}Cl^*$ formed in $^{7}$Be+$^{27}$Al reaction at a temperature T = 3.245 MeV for $ R_a= R_t = R_1 + R_2$ and $ R_a = R_1 + R_2 + \Delta R$, where $\Delta R$ = 1.130 fm i.e. for $\Delta R = 0$ and $\Delta R\neq 0$, respectively. We see that at $\ell = 0\hbar$ the structure of potential energy surface (PES) does not change at these choices. However, we note that at higher $\ell$-values the PES as well as value of fragmentation potential is same except for $^{6}Li$. For the choice of $\Delta R\neq 0$ the fragment $^{6}Li$ has minima. This observation is further reflected in Fig. 3 from the calculated preformation probability $P_0$ using the fragmentation potential given by Fig. 2. At $\ell$ = 0$\hbar$ the LPs are favoured as they have high preformation probability and low fragmentation potential. But at higher $\ell$- values IMFs are energetically more favoured due to low fragmentation potential as shown in Fig. 2 for both the choices of $\Delta R = 0$ and $\Delta R\neq 0$. These results also show that the IMFs $^{6}Li$, $^{8}Be$, $^{10}B$, $^{12}C$, $^{14}N$ are relatively more favoured, having minimum fragmentation potential or higher preformation probability as compared to their neighbors.\\

The summed up preformation probability $\Sigma P_0$ as a function of angular momentum $\ell$($\hbar$), is shown in Fig. 4. At lower $\ell$-value LPs are more favored, but at higher $\ell$-values IMFs starts contributing. As discussed in Fig. 3 the value of $\Delta R$ does not effect the general behaviour of summed up preformation probabilities of LPs and IMFs. However, interestingly there is first increase then decrease in the
$\Sigma P_0$ for LPs just contrary to the behaviour of IMFs with increasing $\ell$-values, for $\Delta R = 0$.\\

Fig. 5 analyze the effects of increasing target mass for $^{7}Be$ induced reactions at $E_{lab} \sim$ 17 MeV, on the fragmentation potential at two extreme $\ell$-values in fig. 5(a) $\ell = 0 \hbar$ and fig. 5(b) $\ell = \ell_{max}$. We see that with increase in mass of the target, the magnitude of fragmentation potential decreases at both $\ell$-values. As we observe in Table. II the temperature decreases with increase in the mass of target. The fragmentation potential depends upon the temperature as given in Eq. 7 (Section II). So the fragmentation potential decreases as the mass of the target increases. The compound systems having higher fragmentation potential have higher decay probability. So, the decay probability of lighter compound system is more, so they have higher value of $\sigma_{fus}$ as compared to the heavier compound systems.\\

We see in Fig. 6 (a) that the barrier increases as we go from light mass region ($^{34}Cl^{*}$) to higher mass region ($^{72}As^{*}$), further imply that the $\sigma_{fus}$ decreases with increase in mass. So we can see in Fig.6(a) fusion cross section decreases with increase in the mass of the target. Fig. 6(b) presents the first and second turning points for $^{34}Cl^{*}$ and $^{72}As^{*}$ decay into $^{6}Li$+$^{28}Si$ and $^{6}Li$+$^{66}Zn$ exit channels, respectively. We see that area under the curve for the case of $^{34}Cl^{*}$ is less as compared to the case of $^{72}As^{*}$, consequently, penetration probability P is more for the former case. It also reason out the large $\sigma_{fus}$ for the compound system $^{34}Cl^{*}$ as compared to the  $^{72}As^{*}$.\\

In Fig. 7 the variation of summed up penetration probability $\Sigma P$  with angular momentum $\ell$($\hbar$) is given for the choices of $\Delta R = 0$ and $\Delta R\neq 0$. Interestingly, at $\Delta R = 0$ $\Sigma P$ for LPs is nearly zero throughout at all the $\ell$-values, but for $\Delta R\neq 0$ the $\Sigma P$ of LPs starts increasing after $\ell$ = $20\hbar$. For IMFs the $\Sigma P$ increases for both the choices of  $\Delta R = 0$ and $\Delta R\neq 0$ at higher $\ell$-values, with the only difference of earlier rise for the later case. The combined results of Fig. 4 and Fig. 7 lead to very small $\sigma_{fus}$ for the choice of $\Delta R = 0$, whereas for $\Delta R\neq 0$ or $\Delta R^{emp}$ $\sigma_{fus}$ are very well compared with the available experimental data and also predicted for some cases as presented in Table II. It clearly indicates the significant role of neck length parameter $\Delta R$ within collective clusterisation description of DCM.\\

To further establish the results of the $^{7}$Be induced reactions, we also investigated the case of other reactions using weakly bound projectiles $^{7}$Li and $^{9}$Be at $E_{Lab}$ = 10 and 28 MeV, for $\Delta R$ = 0.907 and 1.138 fm, respectively, on diffferent targets. The results for the same are presented in Table III and Table IV. Table III presents $\sigma_{fus}$ for $^{7}$Li+$^{27}$Al, $^{7}$Li+$^{28}$Si and $^{7}$Li+$^{59}$Co, compared with the available experimental data. Some predictions are also made by taking stable targets $^{32}S$, $^{40}Ca$ and $^{48}Ti$ at $\Delta R^{emp}$. Table IV gives $\sigma_{fus}$ for $^{9}$Be+$^{28}$Si, $^{9}$Be+$^{124}$Sn and $^{9}$Be+$^{144}$Sm reactions also compared with the available data. The experimental data for the reactions $^{9}$Be+$^{27}$Al, $^{9}$Be+$^{169}$Tm, $^{9}$Be+$^{187}$Re is also available but not for the projectile energy $E_{lab}$ = 28 MeV, but for the neighbouring energies. So, we have extrapolated/intrapolated the data and got $\sigma_{fus}$ which are compared nicely with the results using $\Delta R^{emp}$ calculations.\\

\begin{figure}[t]
\includegraphics[width=0.9\columnwidth,clip=true]{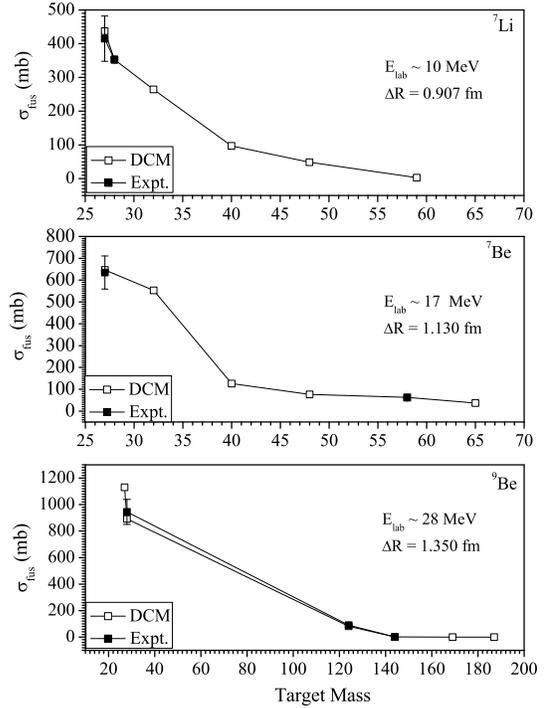}
\vspace{-1.5cm}
\caption{The calculated fusion cross-section $\sigma_{fus}$ for different reactions using different loosely bound projectile and compared with the experimental data}
\label{Fig.8}
\end{figure}

In order to observe the comparative behaviour of $\sigma_{fus}$ of the reactions induced by loosely bound projectiles as a function of target mass, we have plotted Fig. 8. As we have discussed earlier the $\sigma_{fus}$ decreases with increase in the mass of the target. For all the reaction produced by loosely bound projectiles $^{7}Li$, $^{7}Be$, $^{9}Be$, the trend of the fusion cross-section is almost similar. The fitted $\sigma_{fus}$ for a particular projectile at given incident energy find good comparison with the experimental data. The $\sigma_{fus}$ is fitted with same $\Delta R^{emp}$ for particular set of reactions induced by same projectile at the given $E_{lab}$ value. On the basis of these results we have predicted $\sigma_{fus}$ for few reactions induced by these projectile for stable targets with the empirical fitted $\Delta R$ values.The results are depicted in fig 8 and \cite{tab:2}, 3 and 4.

\section{Summary}
The dynamics of reactions induced by loosely bound projectiles has been studied within Dynamical cluster decay model, DCM for a number of reactions. It has been observed that the value of empirically fitted $\Delta R^{emp}$ can be fixed uniquely for particular set of reactions having same projectile and incident energy. Interestingly, for a given loosely bound projectile at fixed incident energy on different targets we are able to calculate/ predict the $\sigma_{fus}$ for all the reactions under study. The results are very well compared with the available experimental data. The calculations with the choices of $\Delta R = 0$ and $\Delta R\neq 0$ clearly points out the significant role played by neck length for the addressal of $\sigma_{fus}$ within the DCM formalism of modified barrier penetration of preformed clusters.Besides the addressing of fusion and predicting cross sections for a number of reactions, an attempt is made to establish the role of target in reaction dynamics. The possible consequence of uniquely fixed neck length parameter is also expressed in context of modification in the barrier characterstics.

In the present study we concentrated only on the total fusion cross section for the given reactions rather than breakup cross sections, complete or incomplete fusion cross sections etc. However, the study on these process is taken up in the future alongwith extension of the present work for the decay of compound nuclei formed in the reactions with a variety of stable projectiles.

\begin{acknowledgments}
BirBikram Singh and Manoj K. Sharma acknowledges the financial support by the Department of Science and Technology (DST), New Delhi under the sanction order SR/FTP/PS-13/2011 and ST/H2/HEP-24/2010 respectively.
\end{acknowledgments}

\end{document}